\documentclass{synmet}
\usepackage{graphicx}

\begin{document}
\begin{frontmatter}
\title{The ubiquitous 1100 charge ordering in organic charge-transfer solids}
\author[label1]{S. Mazumdar,}
\author[label1,label2]{R.~T. Clay,}
\author[label3]{and D.~K. Campbell}
\address[label1]{Department of Physics, University of Arizona, Tucson, AZ, 
	85721, USA}
\address[label2]{ERATO Cooperative Excitation Project,
	Japan Science and Technology Corporation (JST), Tucson, AZ
	85721, USA}
\address[label3]{Departments of Electrical and Computer
	Engineering and Physics, Boston University, Boston, MA 
	02215, USA}

\begin{abstract}
Charge and spin-orderings in the $\frac{1}{4}$-filled organic CT solids are of
strong interest, especially in view of their possible relations to
organic superconductivity.  We show that the charge order (CO) in both
1D and 2D CT solids is of the ...1100... type, in contradiction to
mean field prediction of ...1010... CO.  We present detailed
computations for metal-insulator and magnetic insulator-insulator
transitions in the $\theta$-ET materials. Complete agreement with
experiments in several $\theta$ systems is found.  Similar comparisons
between theory and experiments in TCNQ, TMTTF, TMTSF, and ET materials
prove the ubiquity of this phenomenon.
\end{abstract}

\end{frontmatter}

\begin{figure}
\centerline{\includegraphics[width=2.0in,draft=false]{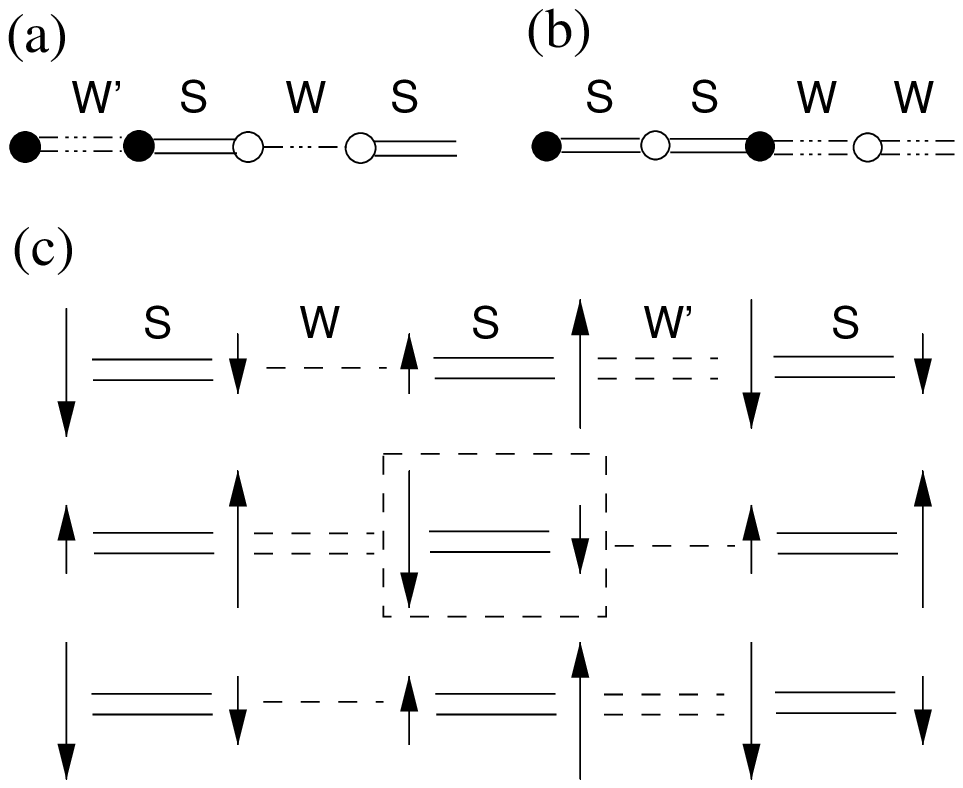}}
\caption{(a) ...1100... CO state. Filled (empty) circles correspond
to sites with more (less) charge. Lines indicate bond order
strength. (b) ...1010... CO state below spin-Peierls transition
(c) The BCSDW in weakly 2D.}
\label{cartoon}
\end{figure}

Ground states involving spatially inhomogeneous charge distributions
have in recent years been observed in a wide range of novel electronic
solids.  In the case of the $\frac{1}{4}$-filled band organic
charge-transfer solids (CTS), it has been suggested that the
appearance of charge-order (CO) may be related to superconductivity
(SC) \cite{Mazumdar00a,Merino01a}.  Two possible CO patterns have been
suggested in 1D systems, corresponding to ``...1010...''  and
``...1100...'' site occupancies (see Figs.~\ref{cartoon}(a) and (b)).
Here `1' and `0' correspond to actual electron or hole densities
0.5+$\epsilon$ and 0.5-$\epsilon$, respectively. The ...1010... CO
corresponds to a 4k$_F$ CDW, where k$_F=\pi/2a$ is the Fermi
wavevector for non-interacting electrons.  In the presence of strong
nearest-neighbor (n-n) Coulomb repulsion $V$, the obvious preferred
state is ...1010... \cite{Seo97a}.  However, this CO can occur only
above a $V>V_{cr}$ (see below).  Furthermore, it is not clear that
this CO can explain the full T-dependent behavior of the
$\frac{1}{4}$-filled band CTS. The ...1100... CO pattern occurs for
$V<V_{cr}$, and is a consequence of {\it cooperative}
electron-electron (e-e) and electron-phonon (e-ph) interactions
\cite{Mazumdar00a}.  This cooperation may be the key element behind
the unconventional SC in the CTS.

Here we examine the theoretical and experimental evidence for both
...1010... and ...1100... CO in the quasi-1D, weakly 2D, and isotropic
2D lattices. We conclude that with the possible exception of
(TMTTF)$_2$X, in all cases the pattern of the CO is ...1100...
Further theoretical and experimental work is necessary to understand
(TMTTF)$_2$X completely.  A second important issue we examine is the
relative temperatures at which CO, metal-insulator (MI), and magnetic
transitions occur.

\begin{figure}
\centerline{\includegraphics[width=3.0in,draft=false]{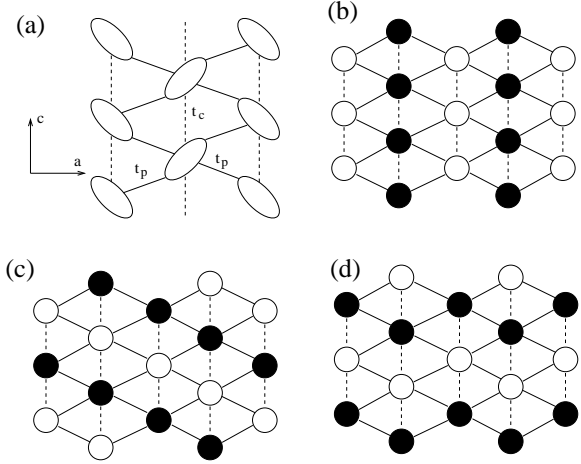}}
\caption{(a) Schematic structure of the $\theta$-(ET)$_2$X conducting
layers, consisting of the ET cationic molecules.  Solid lines
correspond to stronger hopping $t_p$, dashed lines to weaker hopping
$t_c$. (b) vertical stripe, (c) diagonal stripe, (d) horizontal stripe.
Filled (unfilled) circles correspond to molecular sites with greater
(smaller) charge density.}
\label{lattice}
\end{figure}
In 1D, we consider the Peierls extended
Hubbard Hamiltonian, $H=H_0+H_{ee}$, where
\begin{eqnarray}
&H_0&= -\sum_{j\sigma}[t_0-\alpha(u_{j+1}-u_j)][c^\dagger_{j\sigma}
c_{j+1}+h.c] \nonumber \\ &+&\beta \sum_j v_j n_j +
\frac{K_1}{2}\alpha\sum_j(u_{j+1}-u_j)^2 +\frac{K_2}{2}\sum_j v_j^2
\nonumber \\ &H_{ee}& = U \sum_j n_{j\uparrow} n_{j\downarrow}
+V\sum_j n_jn_{j+1} \nonumber
\end{eqnarray}
In $H_0$, $\alpha$ and $\beta$ inter- and on-site e-ph couplings and
$K_1$, $K_2$ the corresponding spring constants. The displacement of
molecular sites from equilibrium is $u_j$, and the on-site phonon
coordinate is $v_j$. $U$ and $V$ are on-site and n-n Coulomb
repulsions.

It is well known that for $\alpha=\beta=0$, the ...1010...  state
occurs only for $V$ larger than some critical value, $V_{cr}$.  For
$V<V_{cr}$, the Luttinger liquid state occurs, which becomes the
...1100... state for nonzero $\alpha$ or $\beta$\cite{Mazumdar00a}.
In the limit $U\rightarrow\infty$, $V_{cr}=2t_0$, and for finite $U$,
$V_{cr}>2t_0$ \cite{Mila93a,Lin95a}.  Consequently, for realistic $U$,
only for systems with the smallest $t_0$ there exists a narrow range
of $V$ for which the ...1010... is a real possibility
\cite{Clay02b}. Since precise estimation of $V$ is difficult, an
alternate approach to determine the CO pattern is to examine the bond
distortion pattern below the spin-Peierls (SP) transition
\cite{Clay02b}. A SP transition within the ...1100... CO has bond
pattern SWSW$^\prime$ (S=strong, W=weak,W$^\prime\neq$W; see
Fig.~\ref{cartoon}(a)), while the bond pattern for the SP phase with
...1010... CO is SSWW \cite{Clay02b}.  Experimental observations of
SWSW$^\prime$ in MEM(TCNQ)$_2$ and other 1D 1:2 TCNQ systems
\cite{Visser83a,Farges85a} provide direct evidence for ...1100... in
the 1:2 anionic CTS. In (TMTTF)$_2$X with smaller $t_0$ (larger
$V/t_0$) CO appears below the MI transition and the situation is more
ambiguous. There exists currently no information on the bond
distortion pattern in the SP phase of (TMTTF)$_2$X.  It has been
suggested that there exists a phase boundary between the CO and SP
phases in (TMTTF)$_2$X \cite{Zamborsky02a}. This might indicate a
switching from ...1010... at high T to ...1100... at low T
\cite{Clay02c}.

In the (TMTSF)$_2$X materials, CO {\it coexists} with SDW order of the
{\it same} (2k$_F$) periodicity \cite{Pouget97a}, which indicates the
coexisting bond-charge-spin density wave (BCSDW) state
(Fig.~\ref{cartoon}(c)) that occurs within a {\it weakly} 2D model
consisting of 1D chains with weak interchain interactions $t_\perp$
\cite{Mazumdar00a}. Evidence for the BCSDW (with very weak moment) is
also seen in $\alpha$-(ET)$_2$X \cite{Mazumdar00a}.

We next consider 2D $\theta$-(ET) lattices (see Fig.~\ref{lattice}).
CO here corresponds to ``stripe'' patterns shown in Fig.~\ref{lattice}
\cite{Seo00a}.  In the limit of $\alpha=\beta=0$ the 2D Hamiltonian
includes weak and strong hoppings $t_c$ and $t_p$ (see
Fig.\ref{lattice}(a)) as well as Coulomb interactions $U$, $V_c$, and
$V_p$. For realistic systems $V_c\sim V_p=V$, and the classical
energies of all three stripes in Fig.\ref{lattice} are equal.  Within
Hartree-Fock (HF) theory\cite{Seo00a}, the vertical stripe state is
found to dominate.

\begin{figure}
\centerline{\includegraphics[width=2.7in,draft=false]{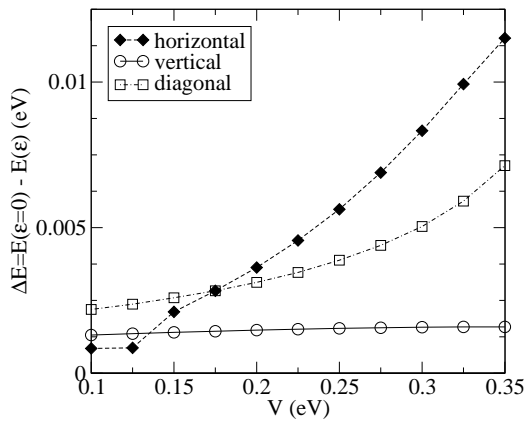}}
\caption{The energy gained upon stripe formation, for the three stripe
patterns of Fig.~\ref{lattice}.  For $V >$ 0.18 eV, the ground state
has the horizontal stripe CO.}
\label{deltae}
\end{figure}

\begin{figure}
\centerline{\includegraphics[width=2.0in,draft=false]{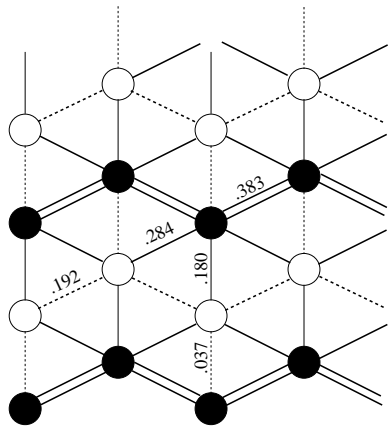}}
\caption{The periodic 16 site lattice investigated for all three
stripe CO orders.  Filled (open) circles indicate large (small) charge
density and numbers are bond orders. See text for parameters. Note the
bond dimerization along the c-direction and the bond tetramerization
along the p-directions.}
\label{thetabow}
\end{figure}

We have performed exact diagonalization studies for a 16-site
$\theta$-lattice with 8 electrons, periodic in all directions (see
Fig.~\ref{thetabow}). Instead of explicitly including the e-ph
interactions as in 1D, we follow our previously established
procedure\cite{Mazumdar00a} and add a site energy component
$\sum_{i}\epsilon_in_i$, where $\epsilon_i$ is negative (positive) for
the `1' (`0') sites. This is equivalent to including $\beta$, with
{\it fixed} $v_j$.  In Fig.~\ref{deltae} we plot the energy gained
upon stripe formation, $\Delta E = E(\epsilon_i=0) - E(\epsilon_i\neq
0)$ for the all three stripes against $V$, for $U$ = 0.7 eV, $t_c$ =
--0.01 eV, and $t_p$ = 0.14 eV, for $|\epsilon_i|$ = 0.01 eV.  We find
that the vertical stripe is never the ground state and the horizontal
stripe dominates for all $V>$ 0.18 eV.  Importantly, while the
vertical and diagonal stripes are ...1010...  along two of the three
directions, the horizontal stripe is ...1100... along the both p
directions (see Fig.~\ref{lattice}).

We also calculate the charge densities $n_j$ and bond orders,
$b_{ij}\equiv\sum_\sigma \langle(c^\dagger_{i\sigma}c_{j\sigma} +
c^\dagger_{j\sigma}c_{i\sigma})\rangle$.  The bond orders are {\it
spontaneously} distorted for the horizontal stripe state in spite of
uniform $t_c$ and $t_p$. This implies spontaneous c-direction
dimerization and p-direction tetramerization for any $\alpha=0^+$ (see
Fig.~\ref{thetabow}). No such bond distortions are found in the
vertical or diagonal stripes.

The above ground state calculations are relevant only at low
temperature. Experimentally, CO at high T ($\sim$190 K)
\cite{Miyagawa00a}, and a spin gap at lower T ($< 50$K) \cite{Mori98c}
have been observed. Transitions at finite temperature are determined
not by the ground state energy, but by the free energy.  At
temperatures T $\leq$ T$_{MI}$, the excitations of a strongly
correlated system are predominantly spin excitations. Because of the
greater multiplicities of high spin states, the free energy at high T
(but below T$_{MI}$) is dominated by high spin states.  We therefore
perform similar calculations of $n_j$ and $b_{ij}$ for the {\it
ferromagnetic} (FM) spin configuration in the horizontal stripe (total
spin S = S$_{max}$ = 4).  The bond dimerization in the c-direction
persists even here, although the p-bonds are dimerized and not
tetramerized. The interpretation of the S = 0 and the S = S$_{max}$
calculations, taken together, is as follows. As T is lowered,
formation of the horizontal stripe CO occurs as soon as charge
excitations are too high in energy and the free energy is dominated by
high spin states.  The CO is accompanied by c-axis bond dimerization,
in complete agreement with experiments \cite{Mori98c}.  At still lower
T, high S states are thermally inaccessible and the free energy is
dominated by the S = 0 state, where there is a {\it second}
transition, also in agreement with experiments. We believe that the
second transition involves tetramerization of the p-bonds.  We have
argued elsewhere \cite{Clay02a} that this change in the bond
distortion pattern from dimerization to tetramerization is accompanied
by a spin gap\cite{Clay02a}. We note that from analysis of $^{13}C$
NMR lineshapes Chiba et. al. have concluded that the CO in
$\theta$-(ET) corresponds to the horizontal stripe \cite{Chiba01a}.

In conclusion, the dominant CO pattern in the $\frac{1}{4}$-filled
band CTS appears to be ...1100... . Direct evidence for this exists in
the 1:2 TCNQ (SWSW$^\prime$ bonds below the SP transition),
(TMTSF)$_2$X (mixed CDW-SDW with same periodicities for CDW and SDW),
and $\theta$-(ET)$_2$X (c-axis bond dimerization at T$_{CO}$ and the
occurrence of two distinct transitions). Further experimental work, in
particular elucidation of the bond distortion pattern below T$_{SP}$
would be necessary to determine the CO in (TMTTF)$_2$X. Most
importantly, the occurrence of the ...1100... CO is a signature of
cooperative e-e and e-ph interactions and this raises the interesting
possibility that the unconventional SC in CTS is a consequence of this
cooperation.

\end{document}